\documentclass[prb,twocolumn,superscriptaddress,nofootinbib]{revtex4-2}

\usepackage{amsmath}
\usepackage{amssymb}
\usepackage{xcolor}
\usepackage{graphicx}
\usepackage{hyperref}
\usepackage{braket}

\begin{document}

\title{Time-domain interferometry of electron weak localization through \\ terahertz nonlinear response}

\date{\today}

\author{Zi-Long Li}
\thanks{These authors contribute equally.}
\affiliation{Institute of Physics, Chinese Academy of Sciences, Beijing 100190, China}
\affiliation{University of Chinese Academy of Sciences, Beijing 100049, China}

\author{Xiao-Hui Li}
\thanks{These authors contribute equally.}
\affiliation{Institute of Physics, Chinese Academy of Sciences, Beijing 100190, China}
\affiliation{University of Chinese Academy of Sciences, Beijing 100049, China}

\author{Yuan Wan}
\email{yuan.wan@iphy.ac.cn}
\affiliation{Institute of Physics, Chinese Academy of Sciences, Beijing 100190, China}
\affiliation{Songshan Lake Materials Laboratory, Dongguan, Guangdong 523808, China}

\begin{abstract}
We study theoretically the nonlinear optical response of disordered electrons in the regime of weak (anti)localization. Our analytical and numerical calculations reveal that, in orthogonal/symplectic class systems, two consecutive, phase coherent optical pulses generates an electric current echo that appears after the second pulse, and at a time equal to the pulse delay time. The current echo reflects the quantum interference between a self-intersecting electron path and its time reversal partner, and, therefore, provide a time-domain interferometry of weak (anti)localization. Our results can be potentially tested on disordered metal films by using terahertz two-dimensional coherent spectroscopy or ultrafast transport measurements.
\end{abstract}

\maketitle

\section{\label{sec:intro} Introduction}

Weak localization is the quintessential \emph{quantum interference} phenomenon that features prominently in two-dimensional disordered conductors~\cite{Bergmann1984,Chakravarty1986,Anderson1979,Gorkov1979}. In such systems, the charge transport is determined by the sum over electron trajectories. Provided that the time reversal symmetry is present and the spin-orbit coupling is weak, a pair of time reversed, self-intersecting trajectories have equal quantum amplitude, and, therefore, would interfere constructively~\cite{Larkin1982}. Electron weak localization emerges from this constructive interference process in that the latter reduces the electrical conductivity from the Drude conductivity. In the opposite limit of strong spin-orbit coupling, the weak antilocalization occurs due to the destructive interference of the trajectory pairs, resulting in an excessive conductivity~\cite{Hikami1980}.

A powerful diagnostic for the electron weak localization is the magnetoresistance~\cite{Bergmann1984,Altshuler1980}. Applying a weak magnetic field perpendicular to the conductor film breaks the time reversal symmetry in a controlled way. The time reversed trajectory pair now pick up Aharonov-Bohm (AB) fluxes that are opposite in sign. The magnetic field suppresses the weak localization by partially destroying the phase coherence between the pair of trajectories. The suppression results in a characteristic magnetoresistance curve, from which key physical quantity such as the electron phase coherence length can be extracted.

The magnetoresistance has been the canonical diagnostic for electron weak localization since the latter's discovery. Yet, other mechanisms for magnetoresistance, such as the Coulomb interaction~\cite{Altshuler1985,Fukuyama1985}, the superconducting fluctuation~\cite{Larkin2005}, as well as their interplay with weak localization, can significantly complicate the analysis of experimental data. Therefore, it is desirable to develop alternative diagnostics, which may allow for a cross examination of the data from independent experimental probes, thereby offering a more comprehensive view on the weak localization phenomenon.

In this work, we address the above problem by proposing that the terahertz nonlinear optical response is up for the task. It has long been recognized that the nonlinear optical response and the quantum interference are deeply linked~\cite{Mukamel1999,Wan2019,Li2021}. We thus anticipate that the nonlinear optical response from disordered electrons may develop unique signatures tied to the weak localization. Meanwhile, the terahertz frequency window matches well the time scale for electron weak localization, namely the phase coherence time $\tau_\phi$, which is on the order of a few picoseconds at a temperature $\sim O(10)$ Kelvin~\cite{Bergmann1984,Lin2002}.

\begin{figure}
\includegraphics[width = \columnwidth]{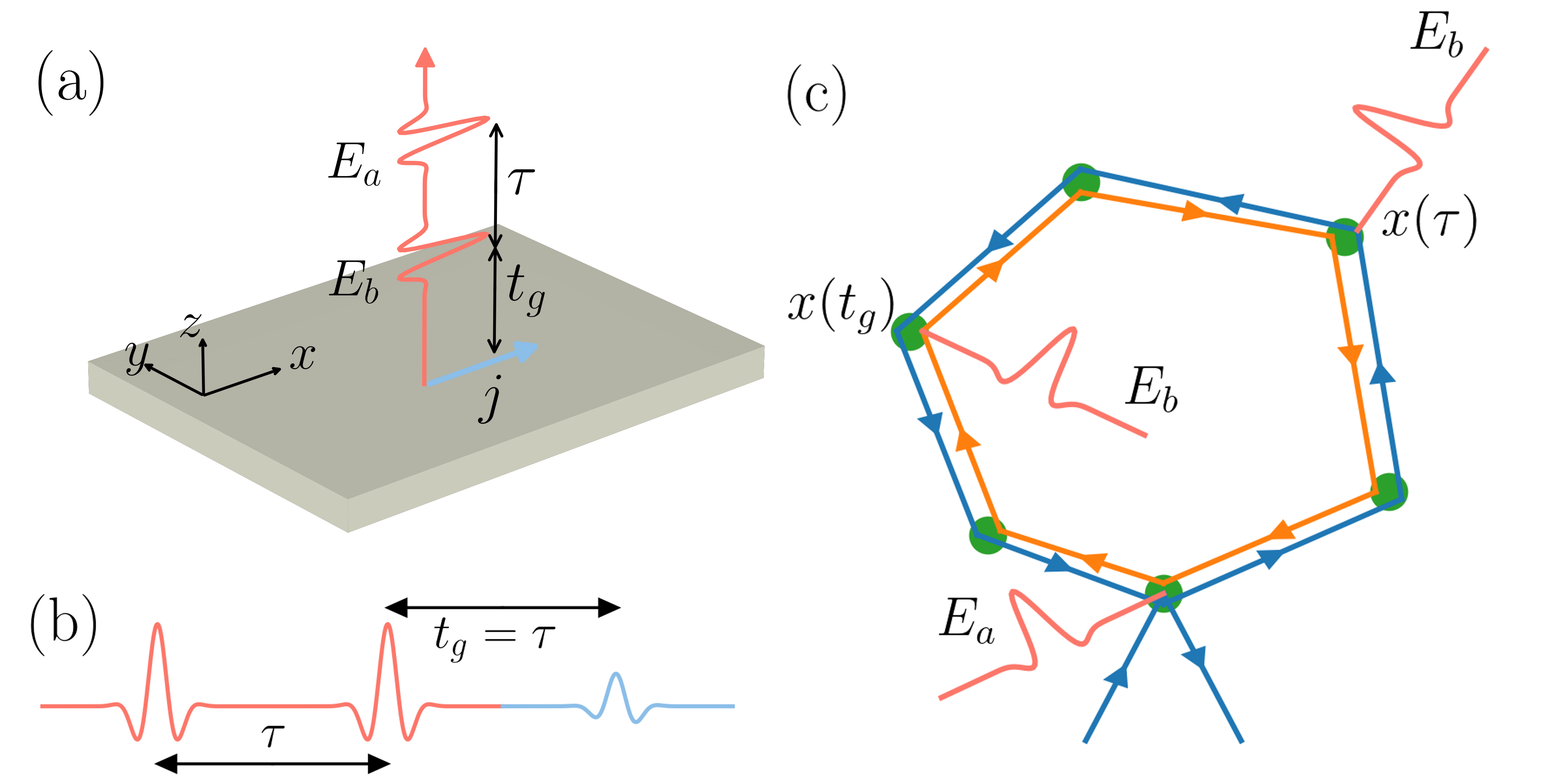}
\caption{(a) Two linearly polarized, phase coherent terahertz pulses ($E_a$ and $E_b$) with delay time $\tau$ generates a nonlinear electric current ($j$) in the disordered metal film. (b) $j$ exhibits an echo at gating time $t_g = \tau$. (c) The current echo arises from the interference between a pair of time-reversed trajectories (blue and yellow). When $t_g\neq\tau$, the trajectory pair acquire different dynamical phases as the electrons are at different locations when $E_b$ kicks in. The loss of phase coherence suppresses weak localization. The weak localization is reinstated when $t_g = \tau$, which produces the echo.}
\label{fig:schematic}
\end{figure}

Specifically, we analyze the nonlinear electric current generated by two consecutive, phase coherent terahertz pulses, with polarization parallel to the conductor film (Fig.~\ref{fig:schematic}a). The two pulses are separated by a delay time $\tau$. For the sake of simplicity, we omit the electron-electron and electron-phonon interactions at the outset. Their impact on weak localization are subsumed in a phenomenological electron phase coherence time $\tau_\phi$ added by hand. We focus on two prototypical symmetry classes of the disorder~\cite{Evers2008}, the orthogonal class and the symplectic class, which describe a disordred metal in the limit of zero and strong spin orbit coupling, respectively. 

Our field theory analysis, as well as numerical simulations, reveal that the nonlinear current, measured as a function of time $t_g$ \emph{after the second pulse} (dubbed the gating time), exhibits a peak (Fig.~\ref{fig:schematic}b). Crucially, the peak appears at the gating time $t_g=\tau$. As the pulse delay time $\tau$ increases, the peak appears later and later. This \emph{echo} behavior originates from the interference of time reversed trajectory pairs, the very same process responsible for weak localization. Heuristically, the first pulse launches electron trajectories at time $0$. Among them, those self-intersect at $t_g+\tau$ contribute to the nonlinear current at the time of measurement. For each trajectory, there is a time reversed trajectory that also contributes. When $t_g\neq\tau$, the electrons are at different positions when the second pulse kicks in. As a result, the time reversed pair would pick up different dynamical phases, thereby suppressing the weak localization. However, when $t_g = \tau$, the weak localization reemerges due to the restoration of time reversal symmetry. This revival of weak localization gives rise to the echo.

The current echo signal, which is also on the picosecond scale, can be potentially detected by the terahertz two-dimensional coherent spectroscopy~\cite{Woerner2013,Lu2018}. In this kind of experiment, one uses two phase coherent terahertz pulses to excite the metal film, and measures the terahertz  electromagnetic field radiated by the nonlinear current. Alternatively, the nonlinear current may be generated and measured on chip by using the latest time-resolved transport measurement techniques~\cite{McIver2020,Wang2023}.

The nonlinear optical response utilizes the delay time $\tau$ and the gating time $t_g$ as knobs to control the coherence between the pair of time reversed trajectories. Therefore, it works as  a \emph{time-domain} interferometry for electron weak localization, which complements the magnetoresistance measurement since the latter, employing the AB effect, can be viewed as a space-domain interferometry. 

Similar to the magnetoresistance, the nonlinear optical response provides a means to probe electron phase decoherence in disordered conductors. Inelastic scatterings due to electron-electron and electron-phonon interactions results in the phase decoherence. Consequently, the current echo fades away exponentially when the pulse delay time $\tau$ increases. In particular, the echo must disappear when $\tau$ is much larger than the electron phase coherence time $\tau_\phi$ as the interference is no longer viable. One may then extract $\tau_\phi$ by carefully monitoring the echo decay. 

The study of nonlinear optical response of disordered electrons has a rich and dynamic history~\cite{Parameswaran2020,Mahmood2021,Paul2022}. It is therefore necessary to put our work in appropriate context. The current echo in the nonlinear optical response of an Anderson insulator, namely the \emph{strong} localization regime, was revealed in Ref.~\onlinecite{Niggemeier1993} by drawing analogy with an ensemble of molecules subject to inhomogeneous environment. The echo mechanism in this case is quite different from the weak localization regime. It is most easily understood in one dimension, where the strong disorder effectively breaks the conductor up into disconnected segments. Each segment can be viewed as a molecule, whose energy spectrum is drawn from a distribution. The echo then arises from the dephasing and rephasing processes triggered by the pulses akin to the Hahn echo~\cite{Hahn1950,Kurnit1964}. Apparently, this mechanism requires no time reversal symmetry since it does not rely on the interference of time reversed electron trajectory pairs.

Much closer to the spirit of this work is Ref.~\onlinecite{Micklitz2015}, where an echo spectroscopy for weak localization was first proposed in the context of cold atoms in optical lattices. It was shown that the breaking and restoration of the time reversal symmetry by a time dependent perturbation can lead to echo phenomenon. Yet, Ref.~\onlinecite{Micklitz2015} focuses on physical observables such as position correlation or momentum distribution, which are natural in cold atoms experiment~\cite{Muller2015} but challenging to access with solid state experimental tools. A main message of the present work, therefore, is that the echo from the \emph{electron} weak localization can be directly observed through ultrafast nonlinear optical response.

The rest of the manuscript is organized as follows. In Sec~\ref{sec:results}, we describe the problem set up and the main results, and provides a quick, heuristic derivation of these results. We give a more rigorous derivation of the results based on the nonlinear sigma model in Sec.~\ref{sec:field_theory}. In Sec.~\ref{sec:numerics}, we test our analytical predictions by numerical simulations. Finally, we discuss issues with experimental feasibility and a few important open questions in Sec.~\ref{sec:discussion}

\section{\label{sec:results} Main results}

In this section, we present the main results from our analytical calculations. We first present a general expression for the electric current response to an arbitrary electric field in Sec.~\ref{sec:nl_current}. We then specialize to the case of two consecutive optical pulses, and analyze the resulted current echo in Sec.~\ref{sec:echo}. Finally, in Sec.~\ref{sec:heuristic}, we give a heuristic derivation of the results.

\subsection{\label{sec:nl_current} Nonlinear current response}

We set the stage by first describing the setup used in this work (Fig.~\ref{fig:schematic}a). It is sufficient for our purpose to consider a dirty metal film consisting of non-interacting electrons. The interaction effects on weak localization are absorbed into a phenomenological constant, the electron phase coherence time $\tau_\phi$. We assume the following hierarchy of time scales, $\hbar/E_F\ll \tau_e \ll \tau_\phi$, where $E_F$ is the Fermi energy and $\tau_e$ the elastic scattering time. Note the time scale of the terahertz pulse is less or comparable with $\tau_\phi$ but much greater than $\tau_e$. In this \emph{diffusive metal} regime, we may focus on the universal behavior of the weak localization as described by the nonlinear sigma model.

We set the metal film in the $xOy$ plane. The film is infinite in both the $x$ and $y$ directions and has zero thickness in the $z$ direction. The linearly polarized, terahertz pulses propagate in $z$ with their polarization $\parallel x$. The electric field, measured on the film, is denoted by $E(t)$. This electric field generates an current, which is $\parallel x$ and uniform in the $xOy$ plane. We denote the sheet current density by $j(t)$.

With this set up, the current density $j(t)$ is given by:
\begin{align}
\label{eq:jt}
j(t) = \sigma_\mathrm{D} E(t) -\eta \frac{G_0}{2\pi} \int^t_{-\infty} \frac{e^{-f(t,t')-\frac{t-t'}{\tau_\phi}}}{t-t'}E(t') dt'.
\end{align}
Here, $\sigma_\mathrm{D}$ is the Drude conductance. The second term describes the weak localization correction. The parameter $\eta $ encodes the underlying symmetry class~\cite{Hikami1980}:
\begin{align}
\eta = \left\{\begin{array}{cc}
1 & (\mathrm{orthogonal}) \\
-\frac{1}{2} & (\mathrm{symplectic})
\end{array}\right. .
\end{align}
$\eta$ reflects the weak localization and antilocalization in the limit of zero and strong spin-orbit coupling, respectively.  $G_0 = 2e^2/h$ is the conductance quantum. 

The coherence factor $f(t,t')$ captures the suppression of the weak localization by a dynamical electric field:
\begin{subequations}
\label{eq:coh_factor}
\begin{align}
f(t,t') = \frac{De^2}{\hbar^2}\int^t_{t'} [A(s) + A(t+t'-s)-2\overline{A}]^2 ds,
\end{align}
where
\begin{align}
\overline{A} = \frac{1}{t-t'} \int^t_{t'} A(s) ds.
\end{align}
\end{subequations}
Here, $A$ is the vector potential in the Coulomb gauge, namely $E = -\partial A/\partial t$.  $\overline{A}$ is the ``moving average" of the vector potential over the time window $(t',t)$. It is easy to check that $f$ is gauge invariant in the sense that shifting $A(t)$ by an arbitrary constant does not change its value. $D$ is the electron diffusion constant of the metal. 

Eq.~\eqref{eq:jt} is applicable to arbitrary electric field $E(t)$ so long as its time scale is much larger than the elastic scattering time $\tau_e$. In the limit of $E(t)\to 0$, $f \to 0$. Eq.~\eqref{eq:jt} reduces to the familiar expression for the weak localization conductivity in the time domain. Said differently, the nonlinear response to a strong electric field is encoded in the coherence factor $f$. Furthermore, the effect of a dynamical electric field is non-perturbative in the sense that expanding $j(t)$ in powers of $E$ yields secular terms. For instance, expanding $\exp(-f)$ to the first order in $f$ results in a third-order nonlinear conductivity $\sigma^{(3)}$, whose magnitude grows with its time arguments, invalidating the naive perturbative expansion in $E$. Crucially, $f$ depends on the temporal profile of $A$ as well as its time reversal with respect to $(t+t')/2$. This structure is responsible for the echo phenomenon described in Sec.~\ref{sec:echo}.

\subsection{\label{sec:echo} Current echo}

\begin{figure}
\includegraphics[width = \columnwidth]{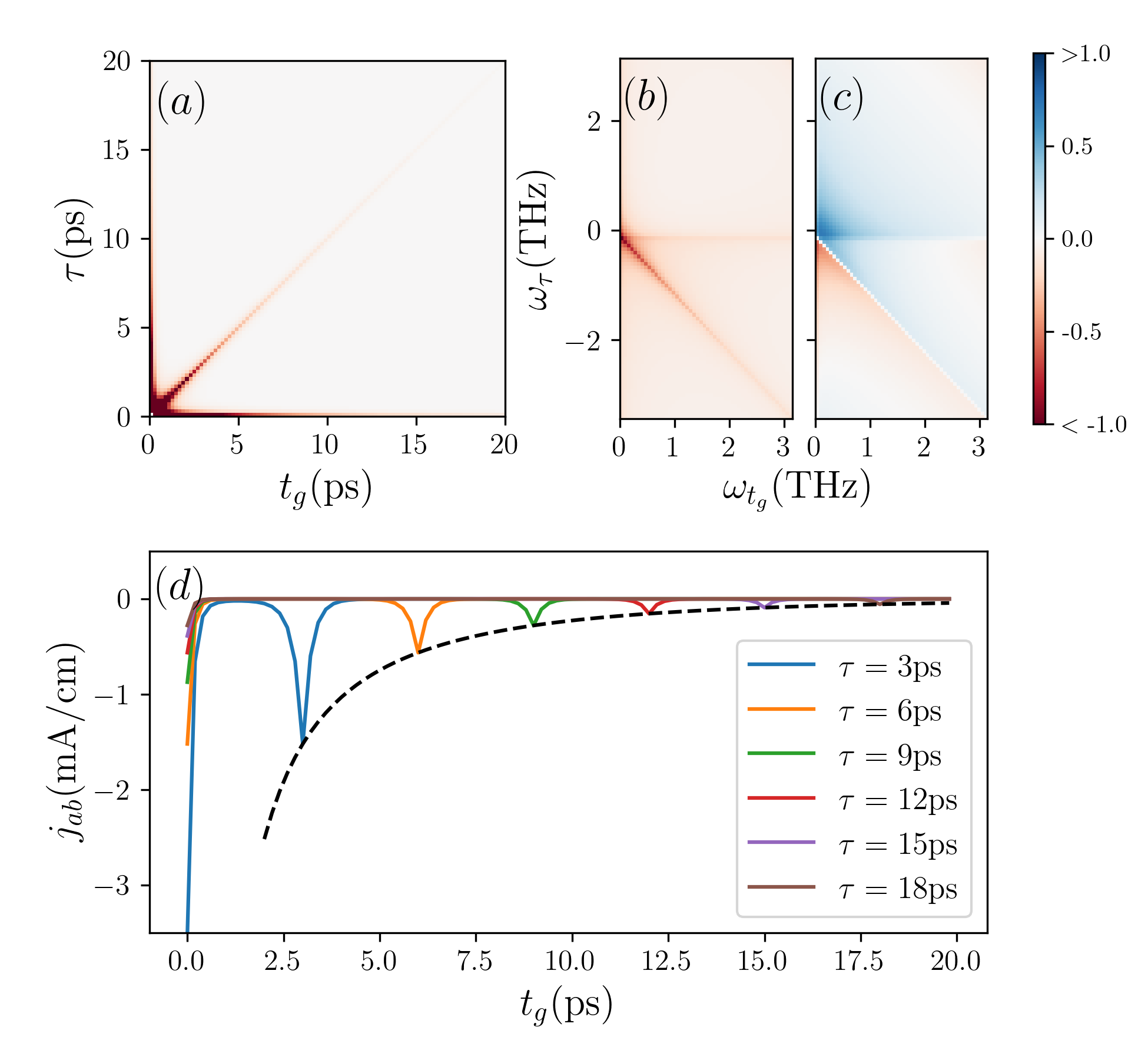}
\caption{(a) Nonlinear current density $j_{ab}$ as a function of the pulse delay time $\tau$ and the gating time $t_g$ for the orthogonal class ($\eta = 1$) and Dirac-$\delta$ pulses. (b)(c) The real and imaginary parts of the two-dimensional coherent spectrum, obtained by Fourier transforming the time domain data shown in (a). Only the first and fourth quadrants are shown as the spectrum in the other half plane is related by complex conjugation. (d) $t_g$ scan of the time domain data for representative values of $\tau$. Dashed line traces the peaks of the current echo. Note the data in (a-c) are in arbitrary units, whereas (d) shows the original data without any rescaling.}
\label{fig:orthogonal_analytic}
\end{figure}

We apply Eq.~\eqref{eq:jt} to the case of two consecutive optical pulses. We illustrate the echo phenomenon by assuming the optical pulses are sufficiently short so that they may be modeled as Dirac-$\delta$ functions:
\begin{align}
\label{eq:E_pulse}
E(t) = E_a\Delta \delta(t) + E_b \Delta \delta(t-\tau).
\end{align}
Here, $E_{a,b}$ and $\Delta$ are respectively the peak electric field strength and the pulse duration. The pulse A arrives at the film at time 0 and the pulse B at time $\tau$. The vector potential is given by:
\begin{align}
A(t) = -E_a \Delta \Theta(t) - E_b\Delta \Theta(t-\tau). 
\end{align}
Substituting the above into Eq.~\eqref{eq:jt}, we find the nonlinear current measured at the time $\tau+t_g$:
\begin{align}
\label{eq:nonlinear_current}
j_{ab}(\tau+t_g)  = -\eta \frac{G_0}{2\pi} \frac{\Delta }{t_g+\tau}e^{-f(\tau+t_g,0)-\frac{\tau+t_g}{\tau_\phi}}E_a.
\end{align}
Here, $j_{ab}$ is the nonlinear current that depends on both $E_a$ and $E_b$, i.e the cross effect of the two pulses. We have dropped terms that depend on $E_a$ or $E_b$ alone. The coherence factor is given by:
\begin{align}
f(\tau+t_g,0) = \frac{2De^2 E^2_b\Delta^2}{\hbar^2}\frac{|\tau-t_g|}{\tau+t_g}\min\{t_g,\tau\}.
\end{align}
$\mathrm{min}\{t_g,\tau\}$ refers to the lesser of the two arguments. 

Fig.~\ref{fig:orthogonal_analytic}c shows $j_{ab}$ calculated from Eq.~\eqref{eq:nonlinear_current} for the orthogonal class ($\eta = 1$). Results for the symplectic class ($\eta = -1/2$) can be obtained  by multiplying the corresponding results by a factor of $-1/2$. We use representative material and pulse parameters: $D = 1$cm$^2$/s, $\tau_\phi = 10$ps, $E_a = E_b = 1$kV/cm, and $\Delta = 1$ps. For fixed pulse delay time $\tau$, we observe a peak in the current response at $t_g = \tau$. This peak is the current echo signal described in Sec.~\ref{sec:intro}. Mathematically, this peak stems from the coherence factor: When $|t_g-\tau|$ is large, $f \gg 1$, which, in turn, suppresses $j_{ab}$ exponentially. However, when $\tau = t_g$, $f=0$, thereby exposing the contribution due to weak localization.

The maximum of $j_{ab}$, located at $t_g = \tau$, is given by:
\begin{align}
j_{ab}(2\tau) = -\eta \frac{G_0}{2\pi} \frac{\Delta}{2\tau}e^{-2\tau/\tau_\phi}E_a.
\end{align}
We see that the height of the peak traces the electron decoherence. As the pulse delay $\tau$ increases, the peak appears later and later; meanwhile, its magnitude decreases. The peak eventually vanishes when $\tau\gg \tau_\phi$.  Thus, we may extract the coherence time $\tau_\phi$ by carefully monitoring how the echo fades away with $\tau$.

The width of the peak $W$ can be tuned by the pulse parameters $E_b$ and $\Delta$, namely $W \propto 1/(E_b\Delta)^2$. Specifically, stronger pulse makes the peak sharper. Experimentally, one may choose appropriate $E_b$ such that the width of the peak matches the time resolution of the instrument.

The echo signal can be measured by using the terahertz two-dimensional coherent spectroscopy~\cite{Woerner2013,Lu2018}. The experimental set up is identical to the one considered in this section. The spectroscopy detects the current echo through the latter's terahertz electromagnetic radiation. Scanning both $t_g$ and $\tau$ produces a two-dimensional plot for the nonlinear signal (Fig.~\ref{fig:orthogonal_analytic}a). The echo manifests itself as the diagonal feature extending up to $\tau_\phi$. The two-dimensional spectrum is then obtained by Fourier transforming the time domain data (Fig.~\ref{fig:orthogonal_analytic}b\&c). The echo appears as a highly anisotropic peak in the fourth quadrant. The width of the peak in the anti-diagonal direction is approximately proportional to $1/\tau_\phi$, whereas the width of the peak in the diagonal direction is controlled by $1/W$.

\begin{figure}
\includegraphics[width = \columnwidth]{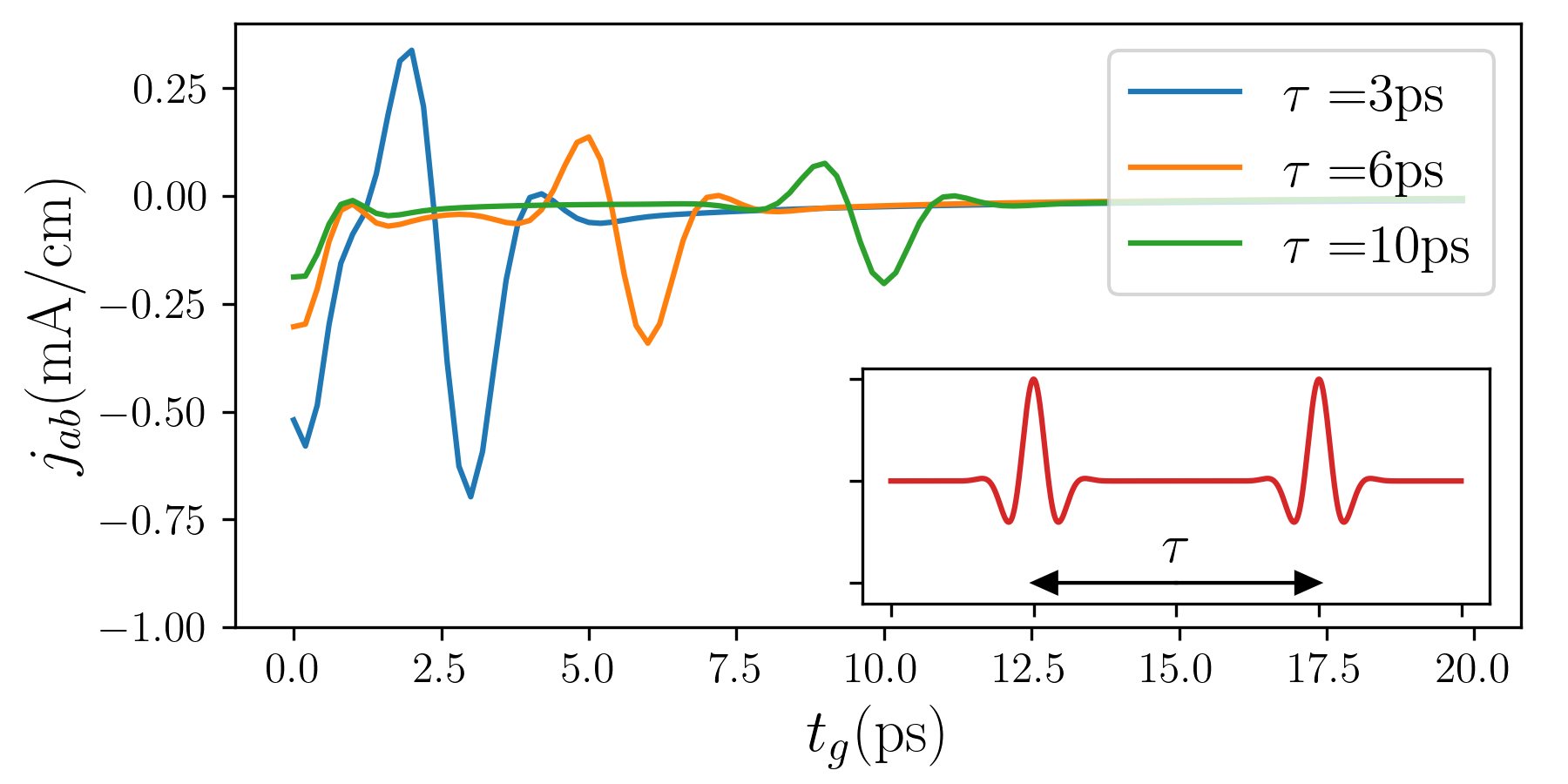}
\caption{Nonlinear current density $j_{ab}$ as a function of gating time $t_g$ (solid lines) induced by single-cycle pulses (inset) for representative values of pulse delay time $\tau$}
\label{fig:realistic_pulse}
\end{figure}

Having illustrated the current echo phenomenon, we now show that the phenomenon is robust with more realistic pulse profiles. To this end, we use single-cycle pulses: $E (t) = E_a e^{-\frac{t^2}{2\Delta^2}}\cos(\omega_0 t) + E_b e^{-\frac{(t-\tau)^2}{2\Delta^2}}\cos[\omega_0 (t-\tau)]$. Fig.~\ref{fig:realistic_pulse} shows the nonlinear current $j_{ab}$ obtained by numerical integration of Eq.~\eqref{eq:jt}. We set the central frequency $\omega_0 = 1$THz. $E_a = E_b = 1$kV/cm. $\Delta = 1$ps. The material parameters are identical to that of Fig.~\ref{fig:orthogonal_analytic}. The current echo peak is clearly visible, but the overall magnitude of the signal is weaker than that of the Dirac-$\delta$ pulses. This is due to the fact that the single-cycle pulse with the same value of $E_a$ and $\Delta$ has less area under the pulse.

\subsection{\label{sec:heuristic} Heuristic derivation}

We derive Eq.~\eqref{eq:jt} heuristically by adapting the semiclassical treatment of Ref.~\cite{Chakravarty1986} to the problem at hand. We consider the orthogonal class ($\eta = 1$) for simplicity. To this end, we consider the probability density for the electron to start at time $t'$ from some position and return to the same position at later time $t$, dubbed $W(t,t')$.  $W(t,t')$ may be expressed as a double path integral:
\begin{align}
W(t,t') = \int Dr' Dr e^{\frac{i}{\hbar}(S[r]-S[r'])},
\end{align}
where $r$ and $r'$ respectively correspond to the forward and backward time evolution. $r(t) = r(t') = 0$, and the same holds for $r'$. $S$ is the action:
\begin{align}
S[r] =\int^t_{t'} [\frac{m\dot{r}^2}{2} - V(r) + e A^\alpha (s)\dot{r}^\alpha (s)] ds,
\end{align}
where $V(r)$ represents the disorder potential. $A^a(s)$ is the time dependent, spatially uniform vector potential due to the terahertz pulse. $\alpha = x,y$ labels the Cartesian components.

Due to the disorder potential, the paths $r$ and $r'$ are phase incoherent except for two special cases. The first case is $r(s) = r'(s)$, i.e. the forward and backward paths are identical. They are always in phase because $S[r] = S[r']$. Responsible for the weak localization is the second case,  $r(s) = r'(t+t'-s)$, i.e. the two trajectories are time reversal partners. We have:
\begin{align}
i(S[r] - S[r']) = ie\int^t_{t'} A^\alpha (s) [\dot{r}^\alpha  (s)  - \dot{r}'^\alpha (s)] ds 
\nonumber\\
 =  ie\int^t_{t'} [A^\alpha (s) + A^\alpha (t+t'-s)]\dot{r}^\alpha (s) ds .
\end{align}
Therefore, they are almost in phase barring the dynamical vector potential. We may express its contribution to $W$ as:
\begin{align}
\widetilde{W} (t,t') = \langle  e^{i\frac{e}{\hbar}\int^t_{t'} [A^\alpha (s) + A^\alpha (t+t'-s)]\dot{r}^\alpha (s) ds} \rangle,
\end{align} 
where the average is over closed paths $r$. $\widetilde{W}$ is a measure of weak localization correction to electron transport~\cite{Chakravarty1986}.

As the electron essentially undergoes random walk due to disorder potential scattering, we may well approximate $r$ as Brownian motion. We discretize the time evolution into slices. Over a time slice $dt$, the electron displacement is given by $dr^\alpha $, where $dr^\alpha $ is a Gaussian random vector with variance $\langle dr^\alpha dr^\beta \rangle = 2Ddt \delta^{\alpha\beta}$, $D$ being the electron diffusion constant. We thus have:
\begin{align}
\widetilde{W} (t,t') = \langle \delta(\sum_s dr_s) e^{i\frac{e}{\hbar}\sum_s (A^\alpha (s)+A^\alpha (t+t'-s))dr^\alpha_s }\rangle.
\end{align}
The average is now over $dr^a_s$. The Dirac-$\delta$ function enforces the constraint that $r$ starts from and returns to the same location. It can be replaced by an integration over the Lagrange multiplier $q^\alpha$:
\begin{align}
\widetilde{W} & (t,t') = \langle \int \frac{d^2q}{(2\pi)^2} e^{i\sum_s [q^\alpha+\frac{e}{\hbar}(A^\alpha (s)+A^\alpha (t+t'-s))]dr^\alpha_s }\rangle
\nonumber\\
& = \int \frac{d^2q}{(2\pi)^2}e^{-D\int^t_{t'} [q^\alpha+\frac{e}{\hbar}(A^\alpha (s)+A^\alpha (t+t'-s))]^2 ds}.
\end{align}
In the second line, we average over $dr^\alpha_s$ and then take the continuous limit $dt\to 0$. Integrating over $q^\alpha$, we obtain:
\begin{align}
\widetilde{W}(t,t') = \frac{e^{-f(t,t')}}{4\pi D(t-t')},
\end{align}
where $f$ is the coherence factor Eq.~\eqref{eq:coh_factor}. Up to a constant prefactor, $\widetilde{W}$ is essentially the same as the weak localization part of Eq.~\eqref{eq:jt}. The phase coherent time $\tau_\phi$ in Eq.~\eqref{eq:jt} is added by hand.

This heuristic derivation shows that the coherence factor captures the suppression of the phase coherence between a pair of time reversed electron trajectories by a time dependent electric field. Specializing to case considered in Sec.~\ref{sec:echo}, we may compare the dynamical phases picked up by $r(s)$ and its time reversal partner $r(\tau+t_g-s)$ (Fig.~\ref{fig:schematic}b)~\cite{Micklitz2015}. For the former path, the dynamical phase due to the pulse B is given by $e\int A^\alpha (s)\dot{r}^\alpha (s)ds /\hbar \sim e{\hbar}\int r^\alpha (s)E^\alpha (s) ds/\hbar = ex (\tau)E_b/\hbar$. For the latter path, the phase is $ex(t_g)E_b/\hbar$. Therefore, the two paths accumulate different phases when $t_g \neq \tau$. However, when $t_g = \tau$, the two paths acquires the same dynamical phase. The loss and reinstatement of phase coherence leads to the suppression and resurgence of the weak localization, which, in turn, produces the current echo.

\section{\label{sec:field_theory} Field theory}

In this section, we justify Eq.~\eqref{eq:jt} by field theory~\cite{Kamenev2011}. Although it can be anticipated from the classic analysis of Ref.~\cite{Altshuler1982}, the present treatment provides a systematic derivation from the perspective of effective action. We illustrate the procedure for the orthogonal class. Since the calculations for the symplectic class are largely in parallel, we refer the interested reader to Appendix~\ref{app:symplectic} for a brief discussion on this class.

In the orthogonal class,  it is sufficient to consider spinless electrons because spin up and down states are decoupled. The starting point is the nonlinear sigma model defined by the action (we use $\hbar=e=1$ in this section):
\begin{align}
\label{eq:NLSM}
iS[Q] = \frac{\pi N_F}{2}\mathrm{Tr}(\check{\partial}_t \check{Q} - \frac{D}{4}\nabla^\alpha_A \check{Q} \nabla^\alpha_A\check{Q}),
\end{align}
$N_F$ is the density of states \emph{per spin} at the Fermi level. $\mathrm{Tr}$ denotes the trace over all the matrix indices as well as integration over space.

$\check{Q}(r)$ is the $4N_t \times 4N_t$ matrix field in the time-reversal $\otimes$ Keldysh space, where $N_t$ is the number of time slices. $\check{Q}$ is subject to the following constraints:
\begin{align}
\check{Q}^2 = \check{I};
\,
\check{Q}^\dagger = \check{Q};
\,
\check{Y}\check{Q}\check{Y} = -\check{Q}^T;
\,
\check{Y} \equiv \begin{pmatrix}
0 & \hat{I} \\
-\hat{I} & 0
\end{pmatrix}.
\end{align}
We use the convention that $\check{M}$ stands for a $4N_t\times 4N_t$ matrix in the time-reversal $\otimes$ Keldysh space, whereas $\hat{M}$ stands for a $2N_t\otimes 2N_t$ matrices in the Keldysh space. Furthermore, $\check{M}_{ts}$ refers to the $4\times 4$ block of $\check{M}$ with designated time arguments $t$ and $s$, whereas $\hat{M}_{ts}$ refers to the $2\times 2$ block of $\hat{M}$.

$\check{\partial}_t$ is the time derivative matrix written in the time-reversal space: $\check{\partial}_t \equiv \mathrm{diag}(\partial_t,\partial_t,-\partial_t,-\partial_t)$. $\nabla^\alpha_A \check{Q}$ is the gauge covariant derivative, $\nabla^\alpha_A \check{Q} \equiv \nabla^a \check{Q} - i[\check{A}^\alpha,\check{Q}]_-$, where $[\cdot,\cdot]_-$ stands for the matrix commutator. $\check{A}^\alpha \equiv \mathrm{diag}(A^\alpha, A^\alpha, -A^\alpha,-A^\alpha)$. We use the Coulomb gauge, $\nabla^\alpha A^\alpha = 0$. 

We seek the effective action $\Gamma[\check{Q}_0]$, where $\check{Q}_0$ stands for the expectation value of $\check{Q}$. This strategy is motivated by the observation that the charge density is related to $\check{Q}_0$ through the following relation:
\begin{align}
\label{eq:charge_density}
\rho(r,t) = \frac{\pi N_F N_s}{4}\mathrm{tr}[\check{\gamma}_q  \check{Q}_{0,tt}(r)].
\end{align}
Here, $N_s = 2$ is the number of spin species. $\check{\gamma}_q$ is the $4\times 4$ charge density vertex: $\check{\gamma}_q = \mathrm{diag}(\hat{\tau}^1, \hat{\tau}^1)$, where $\hat{\tau}^1$ is the first Pauli matrix. As a result, the stationary point of the effective action, $\delta \Gamma[\check{Q}_0]/\delta \check{Q}_0 = 0$, encodes the charge transport equation. In what follows, we illustrate the procedure step by step.

\subsection{Parametrizing the stationary point}

We write the stationary point as:
\begin{subequations}
\label{eq:F-Z}
\begin{align}
\check{Q}_0 (r) = \check{R}(r) \check{\Lambda} \check{R}(r)^{-1}.
\end{align}
$\check{\Lambda} \equiv \mathrm{diag}(I,-I,I,-I)$ is a constant matrix. $\check{R}(r)$ is a block diagonal matrix parametrizes the stationary point:
\begin{align}
\check{R}(r) = \mathrm{diag}[\hat{R}(r), \hat{R}(r)^{T,-1}],
\end{align}
where the block $\hat{R}$, in turn, is given by:
\begin{align}
\hat{R}(r) = \begin{pmatrix}
I & 0 \\
Z(r) & I
\end{pmatrix}\begin{pmatrix}
I & F(r) \\
0 & -I
\end{pmatrix}.
\end{align}
\end{subequations}
$F(r)$ and $Z(r)$ are $N_t\times N_t$ matrix fields parametrizing the stationary point. $F$ plays the role of distribution function, whereas $Z$ describes the deviation of $\check{Q}_0$ from the causal form. When $Z=0$, $\check{Q}_0$ is causal.


We bring in the fluctuations by writing:
\begin{align}
\label{eq:FZ}
\check{Q}(r) = \check{R}(r) \exp(\frac{i}{2}\check{G}(r))\check{\Lambda}\exp(-\frac{i}{2}\check{G}(r)) \check{R}(r).
\end{align}
$\check{G}$ generates soft fluctuations about the stationary point. 
The constrains on $\check{Q}$, as well as the requirement that $\check{G}$ must induces non-trivial rotations on $\check{\Lambda}$, fixes $\check{G}$ to the following form:
\begin{align}
\label{eq:G}
\check{G}(r) = \begin{pmatrix}
0 & d(r) & 0 & c(r) \\
d(r)^\dagger & 0  & c(r)^T & 0 \\
0 & c(r)^\ast & 0 & -d(r)^\ast \\
c(r)^\dagger & 0 & -d(r)^T & 0
\end{pmatrix},
\end{align}
where $c(r)$ and $d(r)$ are $N_t\times N_t$ matrix fields corresponding to the diffuson and Cooperon, respectively.
 
\subsection{Finding the stationary point}

We substitute Eq.~\eqref{eq:FZ} and Eq.~\eqref{eq:G} into the action, and expand it to quadratic order in $c$ and $d$:
\begin{align}
iS[\check{Q}] = iS_0 [F,Z] + iS_1[F,Z,c,d] + iS_2[F,Z,c,d].
\end{align}
$iS_0[F,Z]$ is the value of the action at the stationary point. $S_1$ and $S_2$ are, respectively, linear and quadratic in $c,d$. At one loop, the effective action is given by~\cite{Srednicki2007}:
\begin{align}
\label{eq:effective_action_one_loop}
i\Gamma[F,Z] = iS_0[F,Z] + \log\int DcDd e^{iS_2[F,Z,c,d]}.
\end{align} 
The first term is the tree level contribution; the second term is the one loop correction. 

In principle, we may find the stationary point by first computing $i\Gamma[F,Z]$ and then taking derivatives with respect to $F,Z$. Here, we take a short cut. Setting $Z=0$, the stationary point is causal, and consequently all closed loops vanish. This fact implies $i\Gamma[F,0] = 0$ for any $F$. We deduce that the stationary point is located at $(F,0)$, with $F$ being determined by:
\begin{align}
i \left. \frac{\delta \Gamma}{\delta Z}\right|_{Z=0}  = 0.
\end{align}
Substituting Eq.~\eqref{eq:effective_action_one_loop} into the above equation, we obtain:
\begin{align}
\label{eq:stationary_condition_F}
i \left. \frac{\delta S_0}{\delta Z}\right|_{Z=0} +  \left \langle i\left. \frac{\delta  S_2}{\delta Z} \right|_{Z=0} \right\rangle_{c,d}= 0.
\end{align}
The average is performed with respect to the fluctuations in $c$ and $d$, which are governed by the action $S_2[F,Z=0,c,d]$. 

We now need the explicit expression for $iS_0$ and $iS_2$ to progress further. After some algebra, we find:
\begin{subequations}
\begin{align}
iS_0 = 2 \pi N_F \mathrm{Tr}( [\partial_t,Z]_- F- D(\nabla^\alpha_A F) (\nabla^\alpha_A Z) ).
\\
iS_2 =  -\frac{\pi N_F}{2} \mathrm{Tr}( c^\dagger [\partial_t, c]_+ +  D c^\dagger (-\nabla'_A)^2 c 
\nonumber\\
- 2D(\nabla^\alpha_A Z)c(\nabla^\alpha_AF)^Tc^\ast ) + \cdots.
\end{align}
\end{subequations}
We have dropped from $S_2$ terms that do not contribute to the kinetic equation. $[\cdot,\cdot]_+$ denotes the anti-commutator. ${\nabla'}^\alpha_A \equiv \nabla^\alpha - i[A,\cdot]_+$ is the covariant derivative for the Cooperon field.

Substituting the above expressions into the stationary point condition Eq.~\eqref{eq:stationary_condition_F}, we obtain the kinetic equation, 
\begin{subequations}
\label{eq:kinetic_equation}
\begin{align}
[\partial_t,F(r)]_- + \nabla^\alpha_A J^\alpha_F (r) = 0. 
\end{align}
$J^\alpha_F$ can be interpreted as a current associated to the distribution function:
\begin{align}
(J^\alpha_F (r))_{t_1t'_1}  = -D(\nabla^\alpha_A F(r))_{t_1t'_1} + \frac{D}{\pi N_F} \int dt_2 dt'_2 
\nonumber\\
(\nabla^\alpha_A F(r))_{t_2 t'_2}\mathcal{C}_{t_1 t'_2,t_2 t'_1}(r,r).
\end{align}
\end{subequations}
The first term comes from the classical action $iS_0$; the second arises from the correction due to the fluctuations in the Cooperon field, namely $i\langle \delta S_2/\delta Z\rangle_{c,d}$.  The Cooperon propagator
\begin{align}
\mathcal{C}_{t_1t'_1,t_2t'_2}(r_1,r_2) \equiv  \frac{\pi N_F}{2} \langle c^{\phantom\ast}_{t_1t'_1}(r_1) c^\ast_{t_2t'_2}(r_2)\rangle
\end{align} 
It obeys the Cooperon equation~\cite{Altshuler1982}:
\begin{align}
\label{eq:cooperon}
(\partial_{t_1} - \partial_{t'_1} + D (-i\nabla^\alpha -A^\alpha (r_1,t_1) -A^\alpha (r_1,t'_1))^2
\nonumber\\
 +\frac{1}{\tau_\phi}) \mathcal{C}_{t_1t'_1,t_2 t'_2}(r_1,r_2) = \delta_{t_1t_2} \delta_{s_1s_2}\delta_{r_1r_2}.
\end{align}
Eq.~\eqref{eq:cooperon} can be read off from the kernel of the quadratic action $iS_2$. Here, we have added a mass term $1/\tau_\phi$ by hand to account for electron decoherence effects. 

\subsection{Charge transport equation}

The next step is to extract a charge transport equation from the kinetic equation Eq.~\eqref{eq:kinetic_equation}. It is convenient to perform a change of time variables to the central time $T = (t_1+t'_1)/2$ and the time difference $s = t_1-t'_1$~\cite{Altshuler1982}. The Cooperon propagator is diagonal in $T$:
\begin{align}
\mathcal{C}^{TT'}_{ss'} &\equiv \mathcal{C}_{T+\frac{s}{2},T-\frac{s}{2}; T'+\frac{s'}{2},T'-\frac{s'}{2}} = \mathcal{C}^{T}_{ss'}\delta^{TT'}.
\end{align}
The superscripts (subscripts) correspond to the central times (time differences). The Cooperon equation now acquires a simplified form: 
\begin{align}
 (2\partial_{s} + D (-i\nabla^\alpha -A^{\alpha,T}_s(r))^2  +\frac{1}{\tau_\phi}) \mathcal{C}^T_{ss'} = \delta_{ss'}\delta_{rr'}.
\end{align}
The short hand notation $A^{\alpha,T}_{s}(r) = A^\alpha (r,T+s/2) +A^\alpha (r,T-s/2)$. Meanwhile, the kinetic equation reads:
\begin{subequations}
\begin{align}
\partial_T F^T_s + (\nabla^\alpha_A J^\alpha_F (r))^T_s = 0,
\end{align}
\begin{align} (J^\alpha_F (r))^T_s  = -D(\nabla^\alpha_A F)^T_s  + \frac{2D}{\pi N_F} \int dT' 
\nonumber\\
(\nabla^\alpha_AF(r))^{T'}_s  \mathcal{C}^{\frac{T+T'}{2}}_{T-T'+s,T'-T+s}(r,r).
\end{align}
\end{subequations}

To make contact with the charge density, we substitute Eq.~\eqref{eq:F-Z} (with $Z=0$) into Eq.~\eqref{eq:charge_density}:
\begin{align}
\rho(r,T) = \pi N_FN_s F^T_{s = 0}(r).
\end{align}
Therefore, the charge density is given by the $s = 0$ component of the distribution function $F^T_s$. We observe that $s$ appears as a parameter in the kinetic equation. Setting $s = 0$, and massaging the equations a little, we obtain the charge conservation law, 
\begin{subequations}
\begin{align}
\partial_t \rho + \nabla^\alpha j^\alpha  = 0.
\end{align} 
The electric current $j^a$ obeys a generalized Fick's law:
\label{eq:fick}
\begin{align}
j^\alpha (r,t)   = - \int D_{tt'}(r) (\nabla^\alpha \rho - N_FN_sE^\alpha ) (r,t')  dt'.
\end{align}
The nonlocal diffusion constant is given by:
\begin{align}
D_{tt'}(r) = D \delta_{tt'} - \frac{2D}{\pi N_F}\mathcal{C}^{\frac{t+t'}{2}}_{t-t',t'-t}(r,r).
\end{align}
\end{subequations}
Eq.~\eqref{eq:fick} is the key results of this section. 

Finally, we apply Eq.~\eqref{eq:fick} to the case considered in Sec.~\ref{sec:results}. $\nabla^\alpha \rho = 0$ because  the electric field $E$ is spatially uniform. The electric current reads:
\begin{align}
j (t)  = \sigma_D E (t) - \frac{2N_sD}{\pi}\int \mathcal{C}^{\frac{t+t'}{2}}_{t-t',t'-t}(r,r) E (t')dt'.
\end{align}
$\sigma_D = N_FN_sD$ is the Drude conductance. Solving the Cooperon equation by a spatial Fourier transform, we find the equal-position Cooperon propagator:
\begin{align}
\mathcal{C}^{\frac{t+t'}{2}}_{t-t',t'-t}(r,r) = \theta(t-t')\frac{e^{-(t-t')/\tau_\phi}}{8\pi D(t-t')}e^{-f(t,t')},
\end{align}
where $f(t,t')$ is the coherence factor Eq.~\eqref{eq:coh_factor}. Substituting the above into the expression for the current, we obtain Eq.~\eqref{eq:jt} for the orthogonal class after restoring $\hbar$ and $e$.

\section{\label{sec:numerics} Numerical tests}

In this section, we test the analytic predictions from Sec.~\ref{sec:results} by a direct numerical simulation. We describe our numerical methodology in Sec.~\ref{sec:num_method}. The numerical results are presented in Sec.~\ref{sec:num_results}. Throughout this section, we use the natural units $\hbar = e = 1$. 

\subsection{\label{sec:num_method} Model and method}

We model the disordered electrons by a tight binding model on $L\times L$ square lattice subject to periodic boundary conditions:
\begin{align}
\label{eq:tight_binding}
H = -\sum_{\braket{mn}\sigma} R(m,n)_{\sigma\sigma'}c_{m\sigma}^\dagger c^{\phantom\dagger}_{n\sigma'}   + \sum_{n\sigma} h_n c_{n\sigma}^\dagger c^{\phantom\dagger}_{n\sigma}. 
\end{align}
$c_{n\sigma}$ ($c^{\dagger}_{n\sigma}$) annihilates (creates) an electron with spin $\sigma$ on site $i$. $\braket{mn}$ labels the \emph{oriented} nearest neighbor bonds on the square lattice. We rescale the unit of energy such that the hopping amplitude is $1$. $h_n$ describes the onsite disorder potential; it is drawn uniformly from $[-V,V]$. 

$R(m,n) \equiv R(n,m)^\dagger$ is a $2\times 2$ matrix in the spin space~\cite{Asada2002}. Its form depends on the symmetry class:
\begin{align}
R(m,n) = e^{iA_{mn}(t)} \times \left\{ \begin{array}{cc}
I & \mathrm{(orthogonal)} \\
g\in \mathrm{SU(2)} & \mathrm{(symplectic)}
\end{array} \right. .
\end{align}
$g$ is a random SU(2) matrix drawn uniformly from the Haar measure. $A_{mn}(t) = A^\alpha (t) (r^\alpha_m - r^\alpha_n)$ is the time dependent vector potential on the bond $mn$, through which the electromagnetic field pulse acts on the electrons.

The quantity of interest is the the electric current density generated by the time dependent electric field:
\begin{align}
\label{eq:num_jt}
j^\alpha (t) = \int d\epsilon n_f (\epsilon)\text{tr}(U(t)^\dagger J^\alpha (t)U(t) \delta(\epsilon - H_0)).
\end{align}
$n_f(\epsilon)$ is the Fermi-Dirac distribution function. $J^\alpha (t)$ and $U(t)$ are respectively the current density operator and the time evolution operator in the \emph{one-electron} state space. $H_0$ is the initial Hamiltonian before the impact of the electric field.

We employ the kernel polynomial method to evaluate Eq.~\eqref{eq:num_jt}~\cite{Weisse2006}. To this end, we rescale the Hamiltonian such that its entire spectrum falls inside the interval $[-1,1]$. Expanding the Dirac-$\delta$ function in Eq.~\eqref{eq:num_jt} by Chebyshev polynomials, we obtain:
\begin{align}
\label{eq:num_jt_expansion}
j^\alpha (t) =  \sum^{M}_{n=0} c_n g_n \mu_n(t)   \int d\epsilon \frac{n_f(\epsilon)}{\pi\sqrt{1-\epsilon^2}} T_n(\epsilon).
\end{align}
Here, $M$ is the maximal expansion order. $c_{n} = 1/(1+\delta_{n,0})$. $g_n$ is a coefficient due to the Jackson kernel. $T_n(\epsilon)$ is the $n$-th Chebyshev polynomial of the first kind. $\mu_n$ is the $n$-th Chebyshev moments:
\begin{align}
\mu_n(t) & = \mathrm{Tr}(U(t)^\dagger J^\alpha (t) U(t)T_n(H_0))
\nonumber\\
& = \frac{1}{R}\sum_r \langle r| U(t)^\dagger J^\alpha (t) U(t)T_n(H_0)|r\rangle .
\end{align}
In the second line, we have replaced the trace by an average over random vectors $|r\rangle$, whose components are drawn independently from complex Gaussian distribution with variance $1$. $R$ is the number of random vectors used. The state $|r'\rangle = T_n(H)|r\rangle$ can be computed efficiently using the recursion relation. Meanwhile, we compute the unitary evolution of the state $U(t)|r\rangle$ and $U(t)|r'\rangle$ by using another Chebyshev expansion:
\begin{align}
e^{-iH(t)\epsilon}\ket{\psi} = \sum^{M'}_{n=0} (-i)^n c_n J_n(\epsilon) T_n(H(t))|\psi\rangle.
\end{align}
$|\psi\rangle$ is a state vector. $c_n$ is the same as in Eq.~\eqref{eq:num_jt_expansion}. $J_n(t)$ is the Bessel function of the first kind. $\epsilon$ is a small time interval. The above expansion converges rapidly when $\epsilon$ is small.  The action of $T_n(H(t))$ on $|\psi\rangle$, again, can be computed efficiently by using the recursion relation.

In practice, we use Dirac-$\delta$ pulses with $E_A\Delta = E_B\Delta = 1$ for simplicity. The electric field is polarized along the $x$ direction. The strength of the onsite disorder potential $V = 2$ in the orthogonal class, whereas $V = 1$ in the symplectic class. The system size $L = 1000$. As the Hilbert space dimension is $2\times10^6$, $R=1$ is sufficient for computing the trace. We use $M = 50$ moments for computing the current density and $M' = 10$ for time evolution with $\epsilon = 1$. We average over $500$ disorder samples to obtain good statistics for $j^x(t)$. 

\subsection{\label{sec:num_results} Numerical results}

\begin{figure}
\includegraphics[width = \columnwidth]{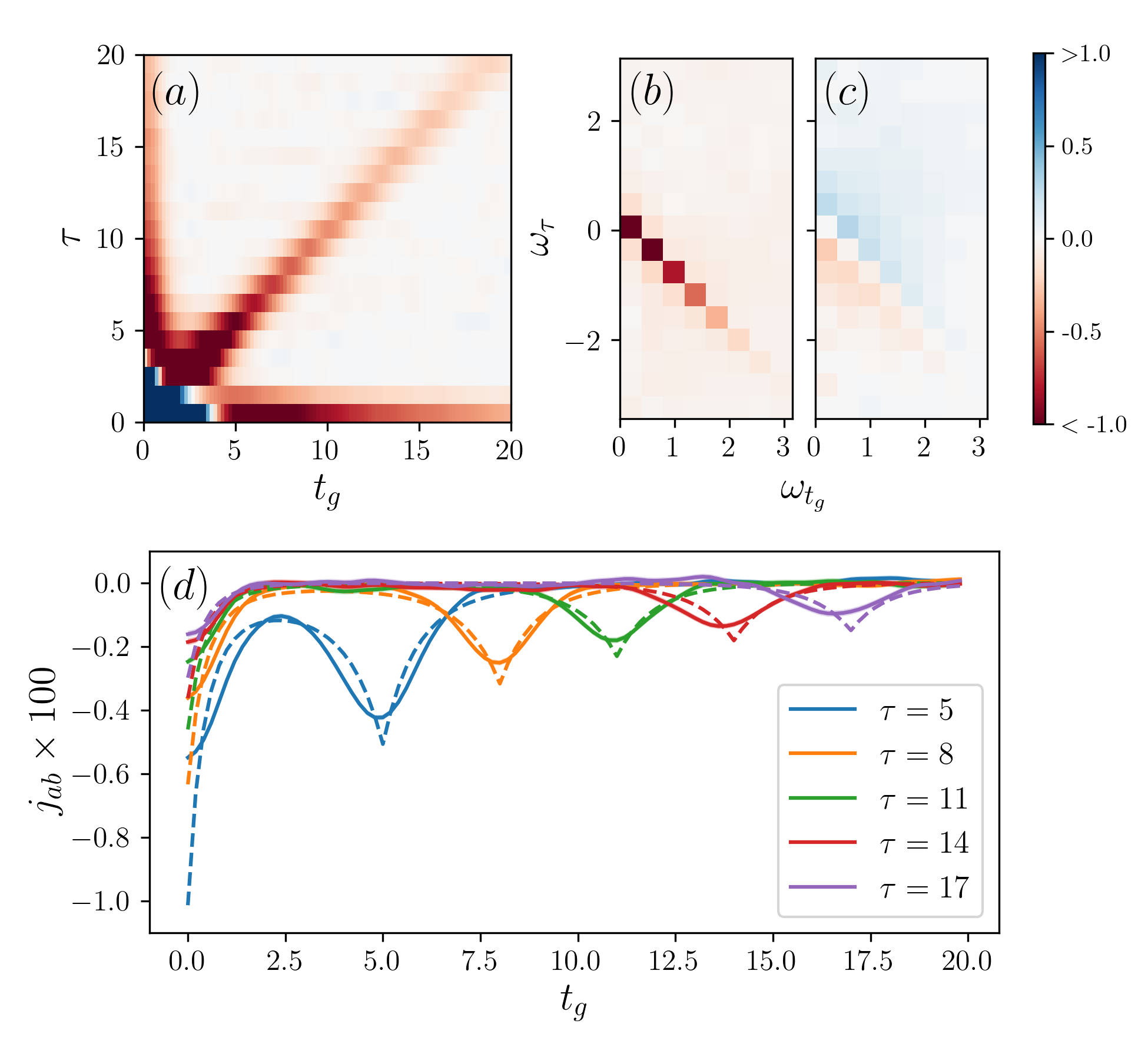}
\caption{(a) Numerically computed current density $j^x(t)$ as a function of the pulse delay time $\tau$ and the gating time $t_g$ in the orthogonal class. The stronger linear response at short times is color saturated in order to reveal the current echo. (b)(c) The real and imaginary parts of two-dimensional Fourier transform of the data in (a).  (d) $t_g$ scan of the data in (a). The solid and dashed lines show respectively the numerical data and the analytical results Eq.~\eqref{eq:nonlinear_current}. The shaded area denotes the error bar.} 
\label{fig:orthogonal}
\end{figure}

Fig.~\ref{fig:orthogonal} shows the numerically computed electric current generated by two consecutive Dirac-$\delta$ pulses. We set the temperature to 0 and the chemical potential $E_F$ to the representative value $-0.6$. As the echo signal is independent of the polarization of the B pulse, we symmetrize our data with respect to $E_b$ and $-E_b$ to remove undesired nonlinear responses. The current echo is clearly seen as the diagonal feature in the two-dimensional plot of the current as a function of the pulse delay time $\tau$ and the gating time $t_g$ (Fig.~\ref{fig:orthogonal}a), in qualitative agreement with the analytical results shown in Fig.~\ref{fig:orthogonal_analytic}. The two-dimensional spectrum also resembles its analytic counterpart (Fig.~\ref{fig:orthogonal}b\&c).

Fig.~\ref{fig:orthogonal}d shows a quantitative comparison between the numerical data (solid lines) and the analytic predictions (dashed lines). To this end, we fit Eq.~\eqref{eq:nonlinear_current} to the numerical data by adjusting the electron diffusion constant $D$, which is the only free parameter. We find $D \approx 1.04$ results in a good fit. We note Eq.~\eqref{eq:nonlinear_current} predicts a cusp in the current echo peak, which is rounded off in the numerical data. This difference is likely due to the details of electronic structure that is not captured by the field theory.

\begin{figure}
\includegraphics[width = \columnwidth]{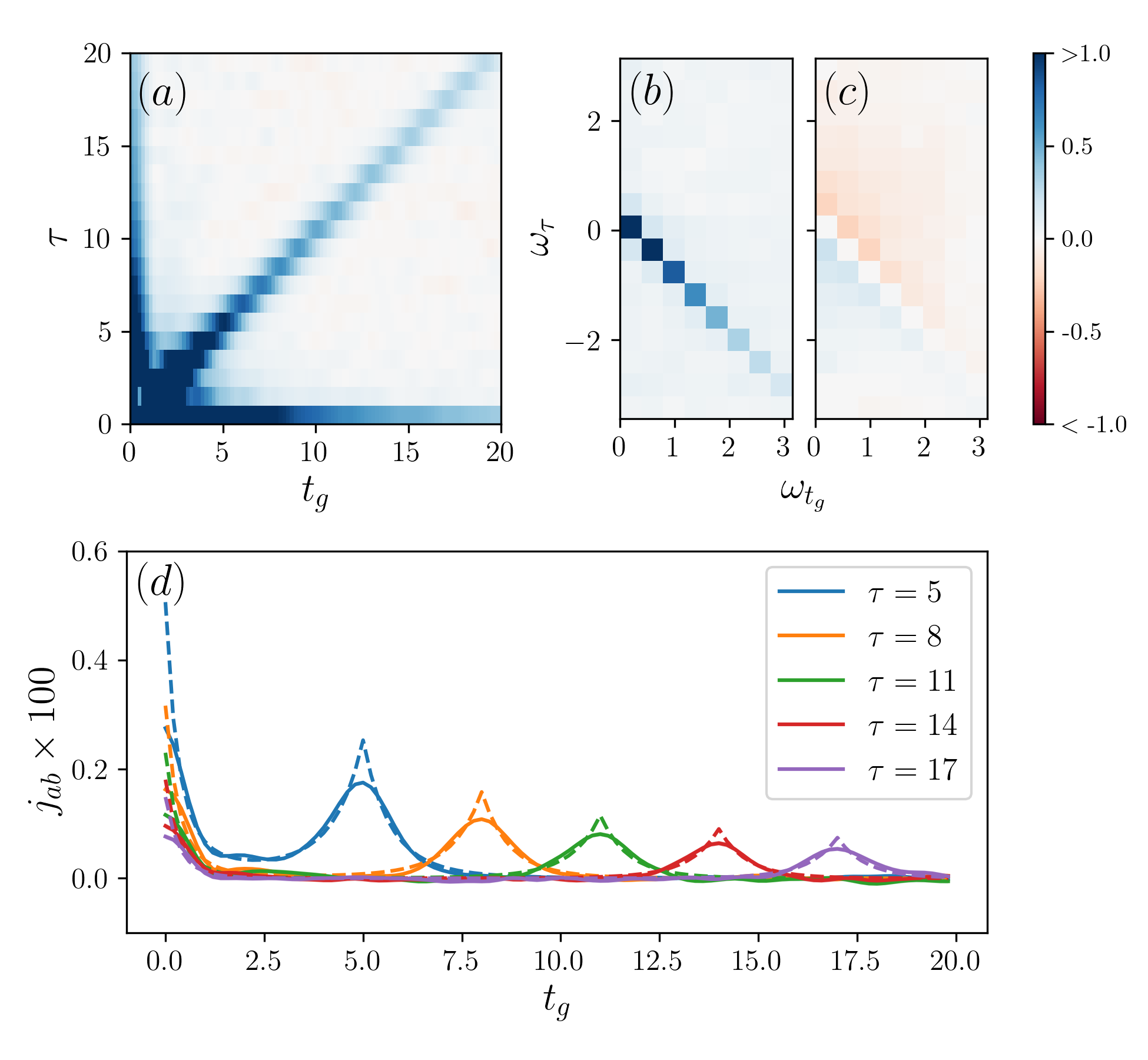}
\caption{Similar to Fig.~\ref{fig:orthogonal} but for the symplectic class.} 
\label{fig:symplectic}
\end{figure}

We find the same good agreement between the numerical data and the analytic predictions in the symplectic class (Fig.~\ref{fig:symplectic}). Here, we set temperature to 0 and the chemical potential $E_F = -0.57$. Compared with the orthogonal class, the echo in the symplectic class reduces by half in its magnitude and changes the sign. A fit of the analytic result Eq.~\eqref{eq:nonlinear_current} to the numerical data indicates that the diffusion constant $D \approx 1.37$.

We further corroborate our analytic results by investigating the behavior of the echo in the presence of an external magnetic field perpendicular to the film. We expect the magnetic field suppresses the echo as it breaks the time reversal symmetry. In the presence of the magnetic field $B$, the nonlinear current reads:
\begin{align}
\label{eq:nonlinear_current_B}
j_{ab}(B,\tau+t_g) = \phi (2BD(\tau+t_g) )\, j_{ab}(0,\tau+t_g),
\end{align}
where $\phi(x) \equiv x/\sinh(x)$. $j_{ab}(0,\tau+t_g)$ is the nonlinear current in the absence of the magnetic field given by Eq.~\eqref{eq:nonlinear_current}. Note Eq.~\eqref{eq:nonlinear_current_B} omits the magnetic field's Zeeman coupling to the electron spin as its impact on the weak localization is negligible compared with the orbital effect. We derive Eq.~\eqref{eq:nonlinear_current_B} by solving the Cooperon equation in the presence of magnetic field, and plugging the resulted Cooperon propagator in Eq.~\eqref{eq:fick}. The details of the calculation is provided in Appendix~\ref{app:magnetic_field}. 

\begin{figure}
\includegraphics[width = \columnwidth]{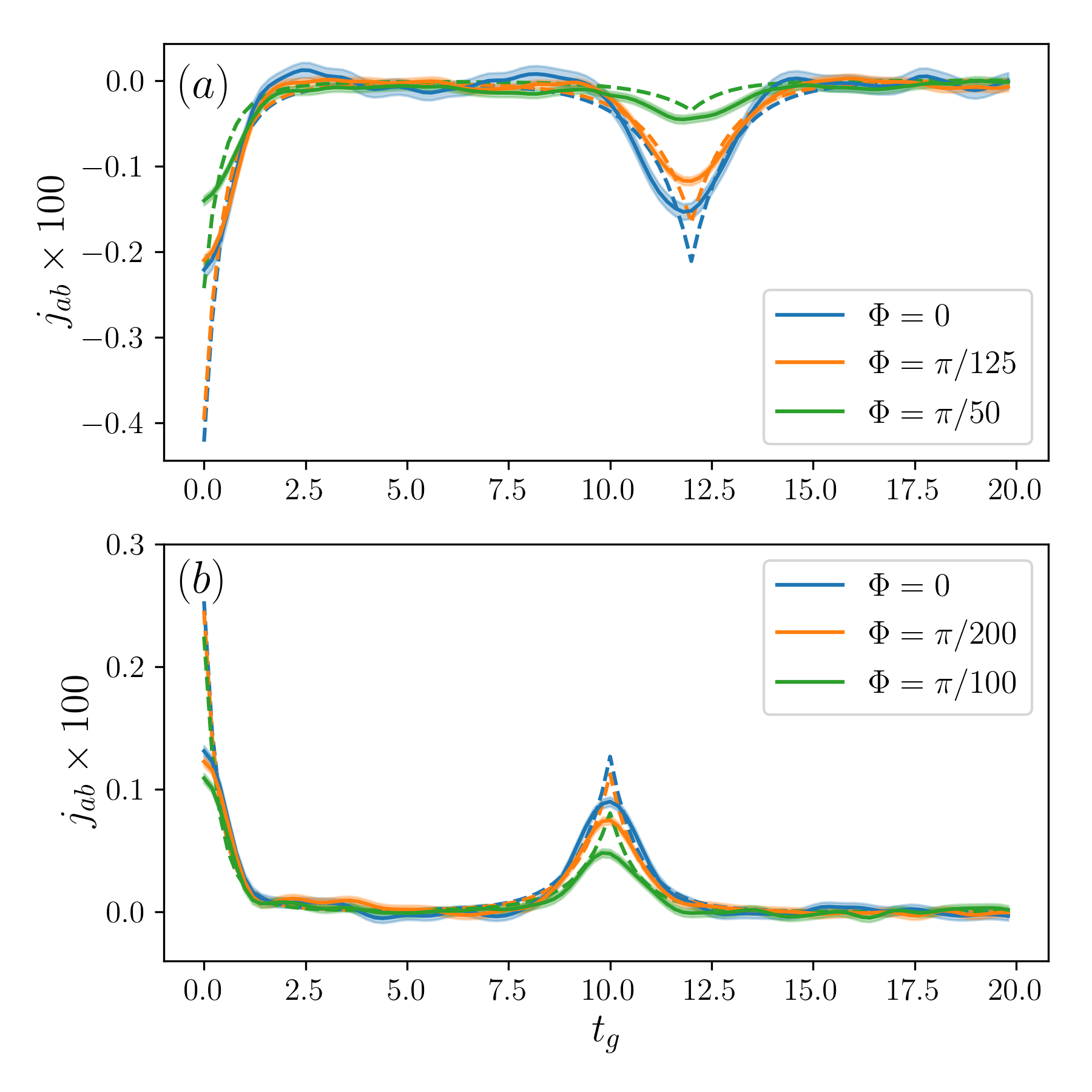}
\caption{The current echo in the presence of a perpendicular magnetic field. (a) The orthogonal class. The solid line represents the numerical data for three representative magnetic field strengths, measured in units of flux per plaquette $\Phi$. The dashed line denotes the analytic result from Eq.~\eqref{eq:nonlinear_current_B}. The shaded area marks the error bar. (b) Similar to (a) but for the symplectic model.} 
\label{fig:current_B}
\end{figure}

Eq.~\eqref{eq:nonlinear_current_B} indicates that the magnetic field results in the suppression of the nonlinear current response. The mechanism is the same as that of the magnetoresistance, namely the time reverse electron trajectory pairs pick up opposite AB phases.  Fig.~\ref{fig:current_B} shows the current echo as a function of the magnetic field, measured in units of the magnetic flux per plaquette, for fixed pulse delay time $\tau$. The model parameters are the same as in those for Fig.~\ref{fig:orthogonal} (orthogonal class) and Fig.~\ref{fig:symplectic} (symplectic class). We see that the echo diminishes as the magnetic field increases for both symmetry classes. Furthermore, we find good agreement between the numerical data (solid lines) and analytic results (dashed lines) with the same fitting parameter, namely $D = 1.04$ for the orthogonal class and $D = 1.37$ for the symplectic class. 

\section{\label{sec:discussion} Discussion}

In summary, we have analyzed the nonlinear optical response of a disordered metal in the weak localization regime. Our main finding is that two consecutive optical pulses, preferably in the terahertz range, can trigger a current echo response. This echo reflects the quantum interference between a pair of time reversed electron trajectories, the same process that produces the weak localization. In particular, one may measure the electron coherence time by carefully monitoring the gradual decay of the echo signal as a function of the pulse delay.

This current echo can be detected by terahertz two-dimensional coherent spectroscopy~\cite{Woerner2013,Lu2018} or ultrafast transport measurement~\cite{McIver2020,Wang2023}, thereby offering an experimental diagnostic for electron weak localization complementary to the magnetoresistance measurement. For a single layer of disordered metal, the current echo signal is on the order of $j\sim 1$mA/cm with an excitation pulse $E_a = 1$kV/cm (Fig.~\ref{fig:orthogonal_analytic}). The electric field strength of the terahertz radiation is on the order of $E=Z_0j/2\sim 0.2$V/cm, which is weak but should be within the reach of available terahertz technology. We expect that stacking multiple layers would enhance the signal strength.

On the theory front, our work leaves a couple of interesting open questions. In deriving our results, we have made the simplifying assumption that the electrons are non-interacting. It is well known that the Coulomb interaction can lead to an anomalous correction to the linear conductivity; this correction has a functional dependence on external magnetic field and frequency similar to the weak localization~\cite{Altshuler1985}. These two contributions must be carefully unravelled in the magnetoresistance measurement. Although we expect that the echo is robust against the Coulomb interaction, the latter might have a nontrivial impact on the nonlinear optical response. The approach presented in Sec.~\ref{sec:field_theory} will prove useful in treating this problem.

Another interesting open problem is the fate of the current echo as the system crossovers from the weak localization to the strong localization. As discussed in Sec.~\ref{sec:intro}, an echo also exists in the strong localization regime, albeit due to a very different mechanism. It is likely that, as the system enters the strong localization regime,  the mechanism discussed in this work gradually subsides as the second mechanism sets in. Exactly in which manner this unfolds is unclear at the moment.

In a broader context, our work attests to the profound connection between the nonlinear response and quantum interference, which we believe produces rich physics that calls for further exploration.

\begin{acknowledgments}
We thank Tao Dong, Chushun Tian, and Yongqing Li for discussions, and Ulrich H\"{o}fer for bringing Ref.~\cite{Niggemeier1993} to our attention. This work is supported by the National Natural Science Foundation of China (Grants No. 12250008 and No. 12188101), by the National Key R\&D Program of China (Grant No. 2022YFA1403800), and by the Chinese Academy of Sciences through the Strategic Priority Research Program (Grant No. XDB33020300) and the Project for Young Scientists in Basic Research (Grant No. YSBR-059).
\end{acknowledgments}

\appendix
\begin{widetext}

\section{\label{app:symplectic} Symplectic class}

The derivation of the charge transport equation in the symplectic symmetry class is in the same vein as the orthogonal class. We therefore focus on the differences between the two classes. In the symplectic case, the spin up and spin down electrons are mixed by the spin orbit coupling. The action of the corresponding nonlinear sigma model reads:
\begin{align}
iS[Q] = \pi N_F \mathrm{Tr}(\check{\partial}_t \check{Q} - \frac{D}{4}\nabla^\alpha_A \check{Q}\nabla^\alpha_A\check{Q}).
\end{align}
Compared with the action for the orthogonal class, there is an extra factor of 2 in front of the action, which reflects the fact that the spin up and down electrons both enter the action. The matrix field $\check{Q}$ is subject to a set of constraints:
\begin{align}
\check{Q}^2 = \check{I};
\,
\check{Q}^\dagger = \check{Q};
\,
\check{X}\check{Q}\check{X} = \check{Q}^\ast;
\,
\check{X} \equiv \begin{pmatrix}
0 & \hat{I} \\
\hat{I} & 0
\end{pmatrix}.
\end{align}
We see that the third constraint is different from the orthogonal class. 

Same as the orthogonal class, we parametrize the stationary point and the fluctuations about it as:
\begin{align}
\check{Q}(r) = \check{R}(r) \exp(i\frac{\check{G}(r)}{2})\check{\Lambda}\exp(-i\frac{\check{G}(r))}{2})\check{R}^{-1}(r).
\end{align}
Here, the definition of $\check{R}$ and $\check{\Lambda}$ are the same as the orthogonal class. In particular, $\check{R}$ parametrizes the stationary point through the matrix fields $F$ and $Z$. However, the matrix field $\check{G}$, which describes fluctuations about the stationary point, acquires a different from due to the different constraint mentioned above:
\begin{align}
\check{G}(r) = \begin{pmatrix}
0 & d(r) & 0 & c(r) \\
d^\dagger(r) & 0 & -c^T(r) & 0\\
0 & -c^\ast(r) & 0 & -d^\ast(r) \\
c^\dagger(r) & 0 & -d^T(r) & 0
\end{pmatrix}.
\end{align}
Exponentiating $\check{G}$ produces the symmetric space $O(4N_t)/[O(2N_t)\otimes O(2N_t)]$. By contrast, exponentiating the $\check{G}$ in the orthogonal class yields the symmetric space $Sp(2N_t)/[Sp(N_t)\otimes Sp(N_t)]$.

Expanding the action up to quadratic order in $c,d$, we obtain:
\begin{subequations}
\begin{align}
iS_0 = 4\pi N_F \mathrm{Tr}[ [\partial_t,Z]F - D(\nabla^\alpha_A F) (\nabla^\alpha_A Z) ],
\end{align}
and
\begin{align}
iS_2 =  -\pi N_F\mathrm{Tr}[c^\dagger [\partial_t, c]_+ + D c^\dagger (-\nabla'_A)^2 c + 2D(\nabla^\alpha_A Z)c(\nabla^\alpha_AF)^Tc^\ast ] + \cdots.
\end{align}
\end{subequations}
We suppress terms in $S_2$ that do not contribute to the kinetic equation. Both expressions gain an extra factor of 2 compared to the corresponding expressions in the orthogonal class. More importantly, the sign in front of the $\nabla^\alpha_A Z (\nabla^\alpha_AF)^T$ term has a plus sign instead of the minus sign. This sign difference is responsible for the weak anti-localization effect in the symplectic class.

The kinetic equation and the charge transport equation can now be derived in the same vein. We find a slightly different Diffusion constant:
\begin{align}
D_{tt'}(r) = D\delta_{tt'} + \frac{D}{\pi N_F}\mathcal{C}^{\frac{t+t'}{2}}_{t-t',t'-t}(r,r).
\end{align}
Here, the propagator $\mathcal{C}$ obeys the same Cooperon equation. We see that the Cooperon correction to the diffusion constant has an extra factor of $\eta = -1/2$ compared to the orthogonal class, which results in the $\eta = -1/2$ factor in the expression for the nonlinear current.

\section{\label{app:magnetic_field} Nonlinear current response in a perpendicular magnetic field}

In this section, we compute the nonlinear current response in the presence of a static magnetic field perpendicular to the film. In the main text, we have shown that the sheet current density induced by a dynamical electric field reads:
\begin{align}
j(t) = \sigma_D E(t) - 4\eta G_0D\int \mathcal{C}^{\frac{t+t'}{2}}_{t-t',t'-t}(r,r) E(t') dt'.
\end{align}
Here, $\sigma_D$ is the Drude conductance. $G_0 = 2e^2/\hbar$ is the conductance quantum. $\eta = 1$ ($-1/2$) in the orthogonal (symplectic) class. Crucially, the weak localization is encoded in the Cooperon propagator $\mathcal{C}$, which is governed by the Cooperon equation:
\begin{align}
[2\partial_s + D(-i\nabla^\alpha - eA^{\alpha,T}_s (r))^2 + \frac{1}{\tau_\phi} ]\mathcal{C}^T_{ss'}(r,r') =\delta_{ss'}\delta^2(r-r'). 
\end{align}
Here, we have used the short hand notation:
\begin{align}
A^{\alpha,T}_s (r) = A^\alpha (r,T+\frac{s}{2}) + A^\alpha (r,T-\frac{s}{2}).
\end{align}
$A^\alpha $ is the vector potential in the Coulomb gauge, $\nabla^\alpha A^\alpha = 0$. 

For the specific case considered here, we write the vector potential as:
\begin{align}
A^\alpha (r,t) = A^\alpha_1(t) + A^\alpha_2(r).
\end{align}
$A^\alpha_1$ describes a spatially uniform, linearly polarized electromagnetic field pulse:
\begin{align}
E^\alpha (t) = -\frac{\partial A^\alpha_1}{\partial t}.
\end{align}
$A^\alpha_2$ describes the static magnetic field perpendicular to the film:
\begin{align}
A^x_2(x) =-\frac{B}{2}y;
\quad
A^y_2(x) = \frac{B}{2}x.
\end{align}

We now need to solve the Cooperon equation for the specific choice of $A$ considered above. The Cooperon equation has the form of the imaginary time Schr\"{o}dinger equation with the Hamiltonian:
\begin{align}
\hat{H}(s) = D(-i\nabla^a-eA^{\alpha,T}_{1,s}-eA^\alpha_2(r))^2 + \frac{1}{\tau_\phi}.
\end{align}
Note $T$ is merely a parameter. As the Hamiltonian at different times $s$ commute, the Cooperon propagator admits the following formal solution:
\begin{align}
\label{app:eq:formal_solution}
\mathcal{C}^T_{ss'}(r,r') = \frac{1}{2}\Theta(s-s') e^{-\frac{s-s'}{2\tau_\phi}}\langle r| e^{-\frac{D}{2} \int^s_{s'} (-i\nabla^\alpha - eA^\alpha_2(r) - eA^{\alpha,T}_{1,u})^2 du }|r'\rangle.
\end{align} 
We simplify the imaginary time evolution operator by expanding the bracket in the Hamiltonian:
\begin{align}
\int^s_{s'} (-i\nabla^\alpha -e A^a_2(r) -e A^{\alpha,T}_{1,u})^2 du & = \int^s_{s'} (-i\nabla^\alpha -eA^\alpha_2(r) -e A^{\alpha,T}_{1,u} + e \overline{A}^\alpha_1 -e \overline{A}^\alpha_1)^2 du
\nonumber\\
& = (-i\nabla^\alpha -e A^\alpha_2(r) - e\overline{A}^\alpha_1)^2 (s-s') + e^2 \int^s_{s'} (A^{\alpha,T}_{1,u} - \overline{A}^\alpha_1 )^2 du. 
\end{align}
Here, $\overline{A}^\alpha_1$ is the temporal average of $A^\alpha_1$:
\begin{align}
\overline{A}^\alpha_1 = \frac{1}{s-s'}\int^s_{s'} A^{\alpha,T}_{1,u}du.
\end{align}
Substituting the above back to Eq.~\eqref{app:eq:formal_solution}, we obtain:
\begin{align}
\mathcal{C}^T_{ss'}(r,r') & = \frac{1}{2}\Theta(s-s') e^{-\frac{e^2D}{2} \int^s_{s'} (A^{\alpha,T}_{1,u} - \overline{A}^\alpha_1 )^2 du}e^{-\frac{s-s'}{2\tau_\phi}}\langle r| e^{-\frac{D}{2} (-i\nabla^\alpha - e A^\alpha_2(r) -e \overline{A}^\alpha_1)^2(s-s') }|r'\rangle
\nonumber\\
 & = \frac{1}{2}\Theta(s-s') e^{-\frac{e^2D}{2} \int^s_{s'} (A^{\alpha,T}_{1,u} - \overline{A}^\alpha_1 )^2 du}e^{-\frac{s-s'}{2\tau_\phi}} e^{ie\overline{A}_1\cdot (r-r')}\langle r| e^{-\frac{D}{2}(-i\nabla^\alpha - e A^\alpha_2(r) )^2(s-s') }|r'\rangle.
\end{align} 
In the second line, we have performed a gauge transformation to remove the constant vector potential $\overline{A}^a_1$ in the Hamiltonian. 

To proceed further, we insert the resolution of identity:
\begin{align}
\mathcal{C}^T_{ss'}(r,r') = \frac{1}{2}\Theta(s-s') e^{-\frac{e^2D}{2} \int^s_{s'} (A^{\alpha,T}_{1,u} - \overline{A}^\alpha_1 )^2 du}e^{-\frac{s-s'}{2\tau_\phi}} e^{ie\overline{A}_1\cdot (r-r')}\sum_\lambda \psi_\lambda (r) \psi^\ast_\lambda (r') e^{-D\lambda (s-s')}.
\end{align} 
Here, $\psi_\lambda$ is the solution for the Landau level problem:
\begin{align}
\frac{1}{2}(-i\nabla^\alpha -eA^\alpha_1(r))^2 \psi_\lambda(x) = \lambda \psi_\lambda(r).
\end{align}
Compared with the standard Landau level problem, the electron mass is $1$, and the magnetic field strength is $2B$. Therefore, the eigenvalues are given by:
\begin{align}
\lambda_{n,m} = (n+\frac{1}{2})2B.
\end{align}
The eigenstates are given by:
\begin{align}
\psi_{n,m}(z) = \frac{1}{2\pi l^2_B}\frac{1}{\sqrt{n!(n+m)!}} (l_B\frac{\partial}{\partial z} - \frac{z^\ast}{4l_B})^n (\frac{z}{l_B})^{n+m} e^{-\frac{|z|^2}{4l^2_B}}.
\end{align}
The magnetic length $l_B = 1/\sqrt{2B}$. $z = x+iy$.

We thus have obtained the general expression for the Cooperon propagator. Now, what enters the expression for the electric current is the equal-space propagator. Owing to the translation invariance of the problem, we may set $r = r'$ to the spatial orgin. We note that $\psi_{n,m}(0) = 0$ when $m>0$. Therefore, the summation is restricted to the subset with $m=0$:
\begin{align}
\mathcal{C}^T_{ss'}(r,r) = \frac{1}{2}\Theta(s-s') e^{-\frac{e^2D}{2} \int^s_{s'} (A^{\alpha,T}_{1,u} - \overline{A}^\alpha_1 )^2 du}e^{-\frac{s-s'}{2\tau_\phi}} \sum_n \psi_{n,0} (0) \psi^\ast_{n,0} (0) e^{-2DB(n+\frac{1}{2}) (s-s')}.
\end{align} 
We now need to evaluate $\psi_{n,0}(0)$:
\begin{align}
\psi_{n,0}(0) = \frac{1}{n!} \frac{1}{2\pi l^2_B}\left. (\frac{\partial}{\partial z} )^n (z)^{n} e^{-\frac{|z|^2}{4l^2_B}}\right|_{z=0}  
= \frac{1}{n!} \frac{1}{2\pi l^2_B}[(\frac{\partial}{\partial z} )^n (z)^{n}] \left. e^{-\frac{|z|^2}{4l^2_B}}\right|_{z=0}  = \frac{n!}{n!}\frac{1}{2\pi l^2_B} = \frac{B}{\pi}.
\end{align}
Substituting the above result back, we obtain:
\begin{align}
\mathcal{C}^T_{ss'}(r,r) &= \frac{1}{2}\Theta(s-s') e^{-\frac{e^2D}{2} \int^s_{s'} (A^{\alpha,T}_{1,u} - \overline{A}^\alpha_1 )^2 du}e^{-\frac{s-s'}{2\tau_\phi}} \frac{B}{\pi}\sum_n e^{-2DB(n+\frac{1}{2}) (s-s')}
\nonumber \\
&= \Theta(s-s') e^{-\frac{e^2D}{2} \int^s_{s'} (A^{\alpha,T}_{1,u} - \overline{A}^\alpha_1 )^2 du }e^{-\frac{s-s'}{2\tau_\phi}} \frac{B}{4\pi\sinh(DB(s-s'))}.
\end{align} 
Setting $T = (t+t')/2$, $s = t-t'$, $s' = t'-t$, the above expression becomes:
\begin{align}
\mathcal{C}^{\frac{t+t'}{2}}_{t-t',t'-t}(x,x) = \Theta(t-t') e^{-f(t,t')} e^{-\frac{t-t'}{\tau_\phi}} \frac{B}{4\pi\sinh(2DB(t-t'))}.
\end{align}
$f(t,t')$ is the coherent factor defined in the main text: 
\begin{align}
f(t,t') & = \frac{e^2D}{2}\int^{t-t'}_{t'-t} (A^\alpha_1 (\frac{t+t'+u}{2}) + A^\alpha_1 (\frac{t+t'-u}{2}) -\overline{A}^\alpha_1) du
\nonumber\\
& = e^2D \int^{t}_{t'} (A^\alpha_1 (u) + A^\alpha_1 (t+t'-u) -\frac{2}{t-t'}\int^t_{t'}A^\alpha (u')du').
\end{align}
Plugging the above result into the expression for the current, we find:
\begin{align}
j(t) =\sigma_D E(t) -\eta \frac{G_0}{2\pi} \int^t_{-\infty} \phi(2DB(t-t')) \frac{e^{-f(t,t')-\frac{t-t'}{\tau_\phi}}}{t-t'} E(t') dt'.
\end{align}
where the function $\phi(x)$ encodes the impact of the magnetic field:
\begin{align}
\phi(x) = \frac{x}{\sinh(x)}
\end{align}
Specializing to the case of two Dirac-$\delta$ pulses, we arrive at:
\begin{align}
j_{ab}(B,\tau+t_g) = \phi(2DB(\tau+t_g)) j_{ab}(0,\tau+t_g),
\end{align}
which is the result given in the main text.
\end{widetext}

\bibliography{disorder_nonlinear}
\end{document}